\documentclass[12pt,preprint]{aastex}

\begin{document}

\title{Bright Localized Near-Infrared Emission at 1--4~AU in the AB~Aurigae Disk Revealed by IOTA Closure Phases}

\author{
R.~Millan-Gabet\altaffilmark{1},
J.~D.~Monnier\altaffilmark{2}, 
J.-P.~Berger\altaffilmark{3},
W.~A.~Traub\altaffilmark{6,4},
F.~P.~Schloerb\altaffilmark{5}, 
E.~Pedretti\altaffilmark{2},
M.~Benisty\altaffilmark{3}, 
N.~P.~Carleton\altaffilmark{4}, 
P.~Haguenauer\altaffilmark{6}, 
P.~Kern\altaffilmark{3}, 
P.~Labeye\altaffilmark{7},
M.~G.~Lacasse\altaffilmark{4}, 
F.~Malbet\altaffilmark{3},
K.~Perraut\altaffilmark{3}, 
M.~Pearlman\altaffilmark{4},
N.~Thureau\altaffilmark{2}}


\altaffiltext{1}{Michelson Science Center, California Institute of Technology, MS 100-22, Pasadena, CA 91125, USA,rafael@ipac.caltech.edu}
\altaffiltext{2}{University of Michigan Astronomy Department, Ann Arbor, MI 48109-1090, USA.}
\altaffiltext{3}{Laboratoire d'Astrophysique de Grenoble, 414 Rue de la Piscine 38400 Saint Martin d'Heres, France}
\altaffiltext{4}{Harvard-Smithsonian Center for Astrophysics, Cambridge, MA 02138, USA}
\altaffiltext{5}{University of Massachusetts at Amherst, Astronomy Department, Amherst, MA 01003, USA}
\altaffiltext{6}{Jet Propulsion Laboratory, California Institute of Technology, Pasadena, CA 91109, USA}
\altaffiltext{7}{LETI-CEA, Grenoble, France}

\begin{abstract}

We report on the detection of localized off-center emission at 1--4~AU
in the circumstellar environment of the young stellar object
AB~Aurigae. We used closure phase measurements in the near-infrared
made at the long baseline interferometer IOTA, the first obtained on a
young stellar object using this technique. When probing sub~AU scales,
all closure phases are close to zero degrees, as expected given the
previously-determined size of the AB~Aurigae inner dust disk.
However, a clear closure phase signal of $-3.5 \pm 0.5 \degr$ is
detected on one triangle containing relatively short baselines,
requiring a high degree of non-point symmetry from emission at larger
(AU-sized) scales in the disk.  We have not identified any alternative
explanation for these closure phase results and demonstrate that a
``disk hot spot'' model can fit our data.  We speculate that such
asymmetric near-infrared emission detected might arise as a result of
localized viscous heating due to a gravitational instability in the AB
Aurigae disk, or to the presence of a close stellar companion or
accreting sub-stellar object.

\end{abstract}


\section{Introduction} \label{intro_sec}

AB~Aurigae (A0~Ve, $d=144$~pc, V~=~7.1, H~=~5.9) is often referred to
as the prototype of the Herbig~Ae/Be (HAeBe) class of pre-main
sequence stars of intermediate mass. Like their solar-type analogs,
the T~Tauri objects, HAeBe objects are known to be surrounded
by pre-planetary disks of gas and dust.  AB~Aur was one of the first
young stellar objects (YSO) to be spatially resolved in the
near-infrared (NIR) \citep{rmg1999}, using the technique of long
baseline optical interferometry. These and subsequent observations of
a relatively large sample of HAeBe (and T~Tauri) objects found
that essentially all objects have characteristic NIR sizes much larger
than expected from previous-generation disk models that were tuned to
fit the spectro-photometric data alone \citep{tuthill1999, rla2000,
rmg2001, rla2002, eisner2004, rla2005, eisner2005, jdm2005}.  These
results have prompted in part a significant revision of disk models,
whereby previously ignored detailed physics of the disk inner edge are
now incorporated with some success \citep{natta2001, dullemond2001,
jdm2002, muzerolle2003,isella2005}. For a recent review of these
developments the reader is also referred to \citet{rmg2006}.

A common characteristic of all the interferometer studies mentioned is
that they were based on single-baseline data, and therefore depended
on the interpretation of only visibility amplitudes and with very
sparse spatial frequency coverage.  One of the unexplored frontiers in
optical long baseline interferometry, particularly in the area of YSO
science, is the measurement and interpretation of not only visibility
amplitudes, but also closure phases (CP), obtained at arrays
containing three or more telescopes. Contrary to the visiblity
amplitudes or phases measured for baseline pairs, the CP is
uncorrupted by atmospheric turbulence \citep{jennison1958,monnier2003},
and can be calibrated to higher precision. The CP can have
values different from 0 or $180 \degr$ only for a non-point-symmetric
sky brightness, and is therefore a highly effective probe of this type
of morphology.  Moreover, CP and visibility amplitude data
may be used together to provide powerful constraints to parametric
models describing relatively complex morphologies. In principle, given
sufficient spatial frequency coverage, direct image reconstruction
becomes possible, as is commonly done at radio wavelengths
\citep{readhead1980,cornwell1981}.

The new models of HAe objects, and of AB~Aur in particular,
successfully reproduce the spectral features and broadband spectral
energy distribution (SED) with a model in which no or little optical
depth exists between the central star and the dust disk inner edge,
such that the frontal stellar heating forms a hot wall of emission
from which essentially all the NIR (disk) flux arises.  This model is
also successful at reproducing the characteristic NIR sizes measured
interferometrically.  Moreover, the outer regions of circumstellar
disks are believed to have scale heights that increase with radius
\citep[flaring,][]{kenyon1987}. Thus, non-point-symmetric brightness
distributions and non-zero CPs are expected for star-disk systems such
as AB~Aur if they have some inclination to the observer's line of
sight, as the cooler foreground disk regions occult part of the hotter
inner disk regions \citep[see also][]{tuthill1999, malbet2001,
isella2005}. The original goal of the work presented here, was to
search for such {\it skewed} emission coming from the inner dust disk;
and in a forthcoming publication \citep{jdm2006} we discuss in detail
the results of our survey of 14 YSOs in the context of hot inner-dust
wall models. In this Letter, we report instead on the unexpected
detection of a strong CP signal that originates further out in the
AB~Aur disk.

\section{Observations and Data Analysis} \label{obs_sec}

The data presented here were obtained at the upgraded three-telescope
Infrared Optical Telescope Array (IOTA) \citep{traub2003}, using the
new integrated-optics (IO) H-band combiner IONIC-3
\citep{berger2003}. IOTA/IONIC-3 has been in routine operation since
December~2002, and first scientific results were published in
\citet{jdm2004} and \citet{kraus2005}.

The three IOTA telescopes are moveable along two orthogonal linear
arms; telescopes A and C can move along a 35~m north-east arm, while
telescope B moves along a 15~m south-east arm. This allows an aperture
of 35~m~$\times$~15~m to be synthesized, giving a maximum physical
baseline $B = 38$~m. For the AB~Aur declination ($\delta = 30 \degr$),
the projected baselines are in the range $B_p = 10 - 38$~m, providing
spatial resolution $\sim \lambda_0/(2 \, B_p) = 4 - 17$~mas (or 0.6 --
2.4~AU) in H-band ($\lambda_0 = 1.65 \mu m$).  The IONIC-3 optical
circuit acts to split the light from each telescope before recombining
each telescope pair (AB, BC, AC) at three IO couplers, leading to six
interferometric channels (two for each baseline).

The work presented here is based on observations made over the period
December~2002 $-$ December~2004, comprising a total of 18~epochs,
14~two-telescope configurations (visibility amplitude data) and
6~three-telescope configurations (CP data).  As is standard practice,
observations of AB~Aur were interspersed with observations of nearby
calibrator stars to measure slowly varying system visibilities and
CPs. We used the calibrator stars HD~32406 (K0III, V~=~6.2, H~=~3.6) and
HD~31233 (K0, V~=~7.2, H~=~4.6); of uniform-disk angular diameters $1.2 \pm
0.7$~mas and $0.6 \pm 0.4$~mas respectively.  The observations
presented here were made using a standard H-band filter ($\lambda_0 =
1.65 \, \mu m$, $\Delta \lambda = 0.30 \, \mu m$) and in un-polarized
light.

Reduction of the visibility amplitude data was carried out using a
custom implementation of the method outlined by \citet{foresto1997}
and already described and demonstrated in \citet{jdm2004}.  To ensure
a good CP measurement, we required that interferograms be detected on
$\geq 2$ baselines, a condition nearly always maintained by a
real-time fringe packet ``tracker" \citep{pedretti2005}.  Based on
studies of our absolute night-to-night calibration accuracy, we adopt
a typical systematic error of $\sigma_{V^2} = 0.05$ (e.g., 2.5\% error
in visibility for unresolved sources); this error has been combined
(in quadrature) with the statistical error for plotting and fitting
purposes.  We followed the method of \citet{baldwin1996} for
calculating the complex triple amplitude in deriving the CP,
explicitly guaranteeing ``fringe frequency closure" ($\nu_{AB} +
\nu_{BC} + \nu_{CA} = 0$). Chromaticity effects are the main
limitation to our absolute precision when the calibrator and source
are of very different spectral types because of different effective
wavelengths. While we minimize these errors by using calibrators of
spectral type relatively similar to that of the target, a conservative
systematic error of $\sigma_{\mbox{CP}} = 0.5 \degr$ has been adopted
here.  Further details on our data reduction and calibration methods
are given in \citet{jdm2006}.  The $V^2$ and CP data have been
averaged in the 2-D spatial frequency $(u,v)$ plane using an averaging
interval of 4~m, resulting in 75 and 37 independent $V^2$ and CP
measurements, respectively. The calibrated $V^2$ and CP data are
available from the authors\footnote{The data can also be found in the
data archives section of the Optical Long Baseline Interferometry News
(OLBIN) web site, {\tt http://www.olbin.jpl.nasa.gov}} in the OI-FITS
format \citep{pauls05}.

\begin{figure}
\begin{center}
\includegraphics[angle=90,scale=0.7]{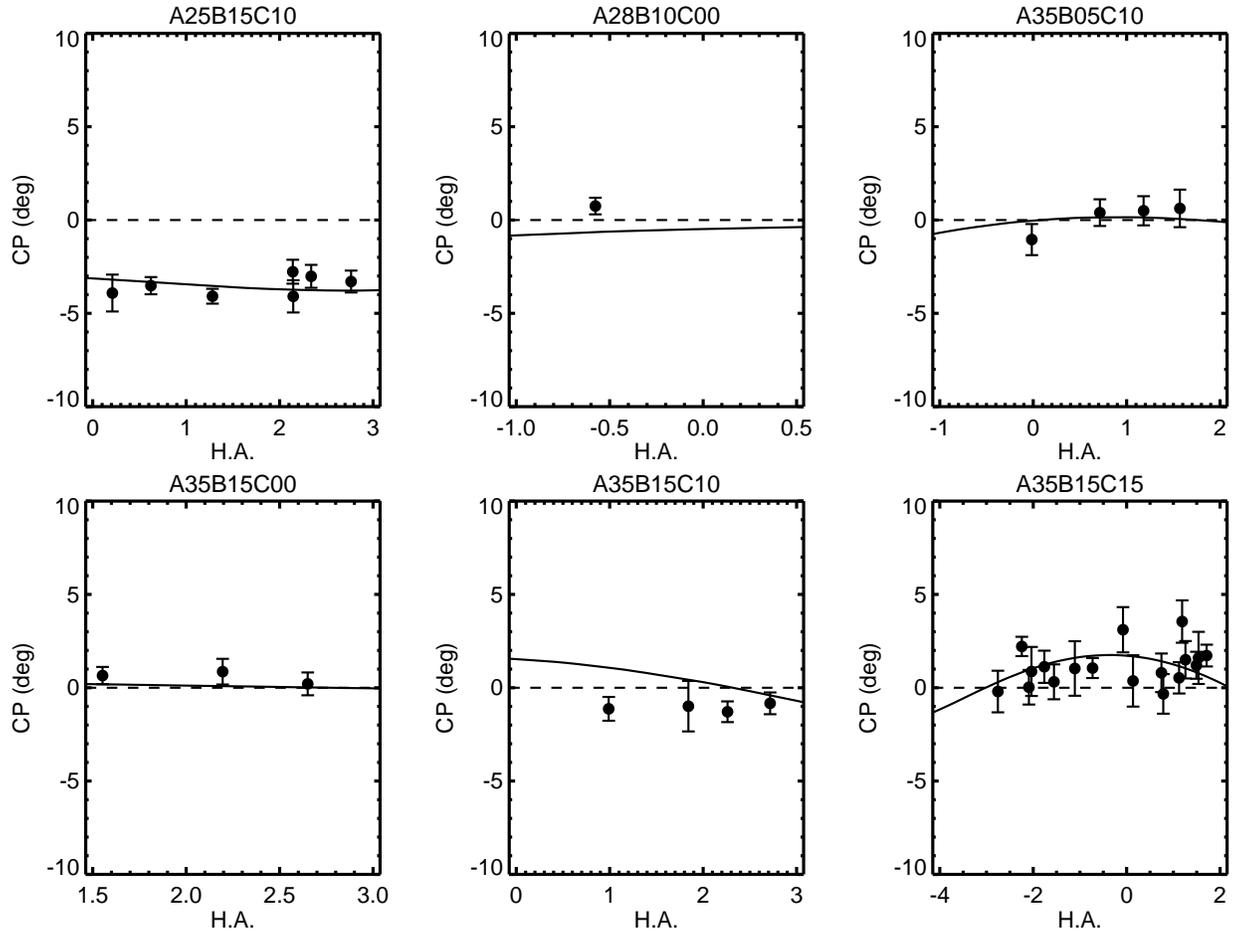}
\caption{Calibrated CP data as a function of hour angle; and comparison
to the models discussed in the text. 
\label{cp_fig}}
\end{center}
\end{figure}

The $V^2$ and CP data are shown in
Figures~\ref{v2a_fig}~and~\ref{cp_fig}, along with models discussed in
the next section.  The main characteristics of our AB~Aur data which
underlie the principal results presented in this Letter are as
follows: (1) A definite CP signal is detected ($-3 \degr$ to $-4
\degr$), which surprisingly appears in a triangle of relatively short
baselines (triangle A25B15C10)\footnote{The notation AxxByyCzz
indicates an IOTA configuration in which telescope A is located at the
station which is xx meters from the array origin, located at the
corner of the L-shaped track, and similarly for telescopes B and C},
indicating some degree of non-point-symmetry on relatively large
spatial scales (1--4~AU, see $\S$~\ref{results_sec}).  (2) The CP data
are consistent with zero on all other baseline triangles
(Figure~\ref{cp_fig}), indicating a high degree of point-symmetry on
sub-AU spatial scales for AB~Aur.  (3) In spite of including baselines
of varying orientations, the curve traced by the $V^2$ {\it vs.}
baseline length data (Figure~\ref{v2a_fig}) is relatively smooth,
indicating that the NIR brightness is also largely circularly
symmetric on small scales (consistent with previous interferometer
observations).  (4) The $V^2$ {\it vs.}  baseline length curve traced
by the data does not appear to extrapolate to 1.0 at zero spatial
frequency, implying the presence of an extended component contributing
a small fraction of the H-band flux ($\sim 8$\% if it were completely
resolved by the shortest IOTA baselines, $B_p = 9$~m). We note that of
the 14 YSOs in our CP survey \citep{jdm2006}, AB~Aur is unique in
showing non-zero CP simultaneous with such high visibilities.

\section{Modelling and Results} \label{results_sec}

The $(u,v)$ coverage obtained in these observations is relatively
dense, unprecedented for YSO observations using an optical
interferometer. However, it still contains significant gaps, a
pronounced north-south bias, and even for the longest IOTA baselines
AB~Aur appears only partially resolved ($V^2_{min} \sim 0.4$); all of
which frustrate attempts to directly reconstruct a meaningful image
from the $V^2$ and CP data.

\begin{figure}
\begin{center}
\includegraphics[angle=0,scale=0.7]{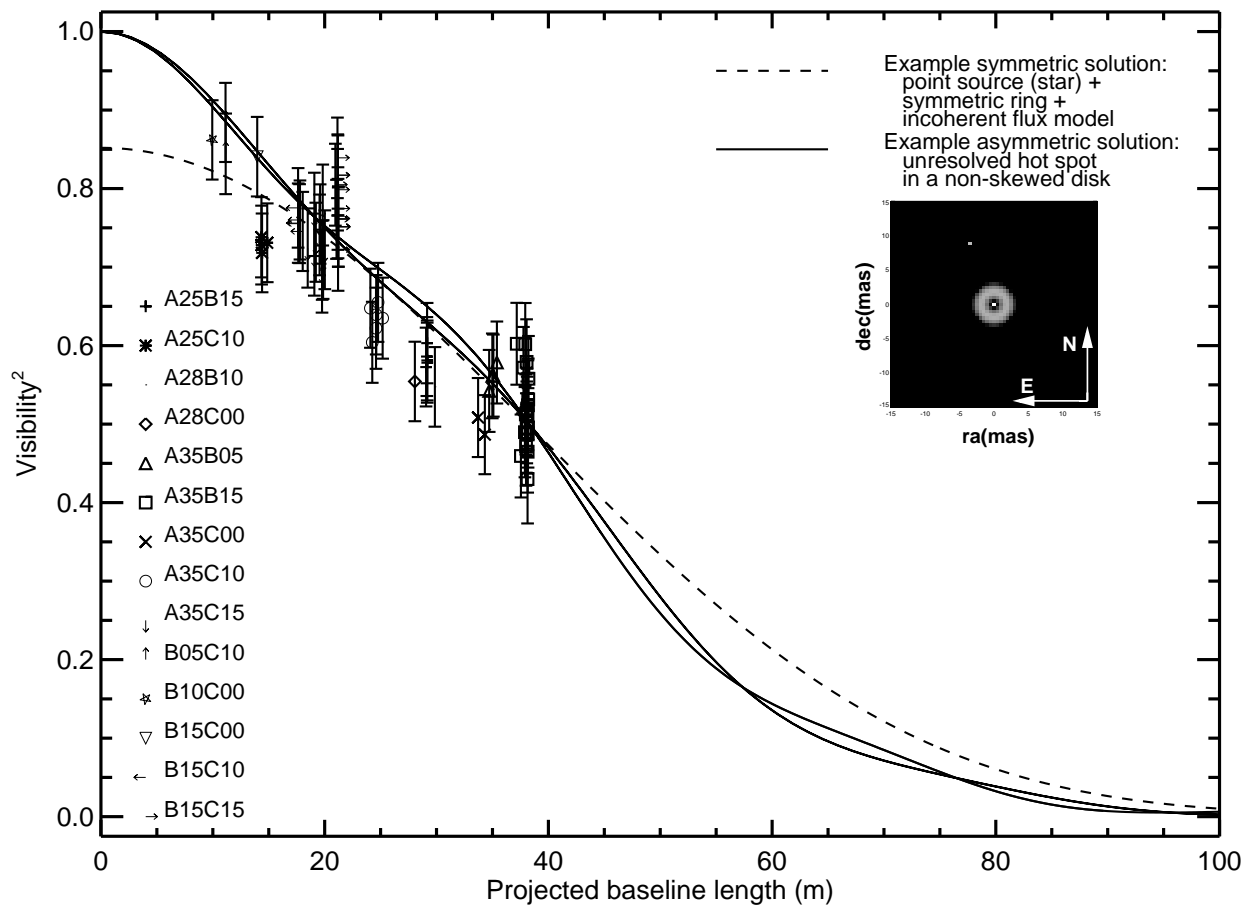}
\caption{
Calibrated $V^2$ data as a function of projected baseline length; and
comparison to the two models discussed in the text.  For the
asymmetric model, the plot shows cuts in the $(uv)$ plane along two
distinct directions that encompass most of the AB~Aur data.  As
discussed in the text, a unique solution for the asymmetric model is
not well constrained; in order to illustrate the geometry the inset
image ($30 \times 30$~mas) shows the example solution for which the
model curves are calculated. 
\label{v2a_fig}}
\end{center}
\end{figure}

We have thus turned to {\it parametric imaging} in order to interpret
our observations. Motivated by the inner dust-wall models introduced
in $\S$~\ref{intro_sec} we begin by modelling the AB~Aur NIR
brightness as central star plus a Gaussian ring, of inner diameter
$D_{ring}$ and width $W_{ring}$ (expressed as fraction of the inner
ring diameter).  We use previous results based on SED decomposition
which have established that the central star in AB~Aur contributes
$\sim 30$\% of the H-band flux and is unresolved (point-like) to the
IOTA baselines \citep{rmg2001} -- the exact stellar fractional flux is
not relevant to the conclusions of this Letter.  Although the
visibility amplitude data allow a direct measurement of the ring
diameter, they do not constrain the ring width well, and we follow
previous workers in adopting a 25\% fractional width for the ring
(also not critical to our conclusions).  Fits of elliptical rings
(representing inclined inner disks) to the $V^2$ data find near-unity
major axis/minor axis ratios, i.e., small disk inclinations $\la 30
\degr$ consistent with previous results from near-IR interferometry
\citep{rmg2001,eisner2004}, scattered light images
\citep{grady1999,fukagawa2004} and mm interferometry
\citep{corder2005}. We therefore adopt a face-on orientation.  From
the best fit star$+$ring model shown in Figure~\ref{v2a_fig} it can be
seen (as mentioned above) that the curve does not extrapolate to $V(0)
= 1.0$, and therefore the model allows for a fraction $f_i$ of the
total flux to be completely incoherent (i.e., large compared to the
angular resolution of the shortest baselines).  The best-fit solution
(to $V^2$ only, resulting in $\chi^2_{\nu}[V^2] = 0.9$) has the
following parameters: $D_{ring} = 3.1 \pm 0.1$~mas
(0.39~$\pm$~0.02~AU), $f_i = 0.08 \pm 0.01$; and is shown with dashed
lines in Figures~\ref{v2a_fig}~and~\ref{cp_fig}. This
value for the ring diameter is in good agreement with recent
interferometer measurements \citep{eisner2004}.

The above model describes some of the main features of the AB~Aur
system (number of components needed, characteristic size and
shape). However, in order to also fit the non-zero CPs, we need to
consider non-symmetric models. As described in $\S$~\ref{intro_sec},
flared disks with non-zero inclination are expected to evince
occultation the near side of the inner disk, resulting in a
non-symmetric brightness. However, as just discussed, the inclination
of the AB~Aur disk is believed to be low.  Moreover, in order to
measure the CP resulting from asymmetries in the inner disk, the
emission has to be well resolved. As shown in \citet{jdm2006} the CP
is strongly suppressed when, as in the AB~Aur case, the inner disk
diameter (3.1~mas) is less than 1/2 the fringe spacing corresponding
to the longest baseline in a triangle.  Here we have confirmed this
result by introducing an azimuthal cosine modulation of the Gaussian
ring in order to approximate the morphology of a partially occulted
inner disk (see \citet{jdm2006} for details of these {\it skewed ring}
models), and found that it can produce at most a signal of
$\mbox{CP[\mbox{A25B15C10}]} \sim -2 \degr$ when the solutions are
constrained to reproduce (even approximately) the $V^2$ data and the
zero CP measured in the other baseline triangles.

We therefore must add an additional source component at larger spatial
scales which introduces additional skewness to the model. In order to
quantify the level of asymmetry that is required, we have adopted a
simple Gaussian profile which is offset with respect to 
the central disk emission (``Disk Hot Spot'' model). 
The new component is thus characterized by its full-width at
half-maximum ($W_G$), location (angular separation $r_G$ and position
angle $\mbox{PA}_G$) and the fractional flux ($f_G$) it
contributes. Fitting this simple model we find that solutions exist
that reproduce both the $V^2$ and CP data reasonably well. However, we
emphasize that several solutions exist which are statistically
equivalent, and although we can assign definite spatial scales and
fractional fluxes to the offset Gaussian, we are not able to assign a
definite direction to the asymmetry (e.g., there are reasonable solutions
with northern as well as southern offsets).  We have included
the best-fit parameters associated with a few of the best solutions
in Table~\ref{fits_tab} for reference.

Quantitatively, the solutions found are defined by the following
characteristics: (1) when the offset Gaussian feature is allowed to be
spatially extended, the angular separations are $r_G \simeq 30$~mas
(4.3~AU), width $W_G \simeq 12$~mas (1.7~AU) and fractional flux $f_G
\simeq 8$\% (we also note that in this case slightly better fits are
found when the inner ring is also allowed to be skewed, rather than
symmetric); and (2) for the special case that the offset component is
unresolved (i.e., a point source companion), closer separations $r_G
\simeq 9$~mas (1.25~AU) and lower fractional fluxes $f_G \simeq 2$\%
are found (in this case with no benefit from additional skew in the
inner ring). An example fit of the latter case is shown (solid lines)
in Figures~\ref{v2a_fig}~and~\ref{cp_fig}. The example
shown is typical of the quality of the fits obtained
($\chi^2_{\nu}(\mbox{[$V^2$,CP]} = 1.5$) and corresponds to the
parameters: $f_G = 0.02$, $r_G = 9$~mas (1.4~AU) and $\mbox{PA}_G = 22
\degr$.  As can be seen, although the solution is not entirely
adequate on all baseline pairs and triangles, indicating the need for
a more complex model, this simple model is capable of reproducing most
data well, in particular the observed $\mbox{CP[\mbox{A25B15C10}]} =
-4 \degr$.

\section{Discussion}

The principal result of this Letter is the clear detection of an
unexpected localized asymmetry in the circumstellar environment of
AB~Aur at spatial scales of 1--4~AU; this is an {\it intermediate} scale
between the the inner wall containing the hottest dust and the outer
disk.

We first considered that the physical origin of this emission could be
H-band scattering off the disk atmosphere or ``halo''.  The required
asymmetry could easily arise from a small photocenter shift due to
e.g., forward scattering in combination with a small disk
inclination. However, this emission would arise from large angular
scales (up to $\sim 1 \arcsec$) and would be completely resolved by
even the shortest IOTA baselines, and therefore could not result in a
CP signal.  Indeed, the model fitting described above shows that the
offset Gaussian must be relatively compact ($\lesssim 12$~mas or
1.7~AU), and it is difficult to reconcile the required amount of flux
with scattering over such a small area.  Due to the low inclination
known for the AB~Aur disk, we also do not favor outer disk occultation
or opacity effects.

A more likely explanation, is that the asymmetry is due to compact
thermal emission. The presence and persistence of such a local
inhomogeneity in the AB~Aur disk at these spatial scales would however
be surprising, given that the dynamical time-scales are very short
(1--5~years) compared to the age of the system ($\sim 1-4$~Myr).  A
possible explanation is that the disk is not steady state, and the
feature corresponds to a gravitational instability, possibly
associated with an inner extension of the spiral density waves seen in
near-IR coronographic and millimiter interferometry images
\citep{fukagawa2004,pietu2005}.  The existence of a disk instability
at few~AU radii and for a disk of this age clearly has important
implications for the formation of planets in these disks.

Alternatively, we could be detecting a companion embedded in the
AB~Aur disk.\footnote{We note that \citet{baines2006} report
``spectro-astrometric'' evidence for binarity in AB~Aur (and several
other HAeBe objects). However, they infer large separations of $0.5 -
3.0 \arcsec$; therefore this putative companion can not be related to
the CP signatures modelled here.} A stellar companion at the few--AU
separation we infer is however unlikely, given that a substantial
circumstellar disk has in fact survived around AB~Aur.  The intriguing
possibility therefore arises that the companion is instead a
proto-planetary object forming in the disk.

Similar claims (but based on visibility amplitudes rather than the
more robust CP) for a disk thermal instability or close companion have
been recently made for the active-disk system FU~Orionis
\citep{malbet2005}. We also note the possible connection between the
extended structure detected here, and the intriguing compact halo
(45~mas FWHM) seen in the T~Tauri object DG~Tau in lunar occultation
observations by \citep{leinert1991}. As discussed by \citet{jdm2006},
disk ``halos'' appear to be common (although not the norm) 
in young star~$+$~disk systems, and deserve new observational and modelling scrutiny.

Our hypothesis is readily testable. If the non-zero CP arises in a
``disk hot spot,'' variability in the CP signal over 1--5~years should
reveal the orbital motion that could be followed using short or long
baselines from the VLTI AMBER \citep{malbet2004} or MIDI
\citep{leinert2003} instruments, respectively.  The angular scales
involved also make this object a prime target for direct imaging using
high resolution techniques on large single-aperture telescopes.

\acknowledgments

The authors gratefully acknowledge support from SAO, NASA (for third
telescope development \& NNG05G1180G), the NSF (AST-0138303,
AST-0352723, for work on imaging with IOTA), and the Jet Propulsion
Laboratory (JPL awards 1236050 \& 1248252).  The IONIC3 instrument has
been developed by LAOG and LETI in the context of the IONIC
collaboration (LAOG, IMEP, LETI). The IONIC project is funded by the
CNRS (France) and CNES (France).

\clearpage
\begin{deluxetable}{lcccllc}
\tabletypesize{\tiny}
\tablecolumns{6}
\tablewidth{0pc}
\tablecaption{Results from Fitting to ``Disk Hot Spot'' Model\tablenotemark{a} \label{fits_tab}}
\tablehead{
\colhead{Model} & \multicolumn{3}{c}{Fraction of Light} & \colhead{Disk Properties} & \colhead{Spot Properties} & \colhead{Reduced $\chi^2$}\\
\colhead{Description} & \colhead{Star} & \colhead{Disk} & \colhead{Spot} & & & \colhead{(V$^2$,CP)}
}
\startdata
Unresolved hot spot   & 0.3 & 0.68 & 0.02 & Ring Diameter 3.6~mas    & Unresolved Spot & 1.5\\
with non-skewed disk\tablenotemark{b} &     &      &      & Ring Width/Diameter 0.25 & $r_G=$9~mas at PA 22$\arcdeg$ &\\ 
\hline
Gaussian hot spot     & 0.3 & 0.62 & 0.08 & Ring Diameter 3.1~mas    & Gaussian FWHM 12~mas & 1.8 \\
with skewed disk    &     &      &      & Ring Width/Diameter 0.5  & $r_G=$29~mas at PA 12$\arcdeg$ & \\
                  &     &      &      & Max Skew$=$1.0 at PA 172$\arcdeg$ & \\
\enddata
\tablenotetext{a}{See description of all parameters in text (\S\ref{results_sec}).  
Error bars are not presented in this table since multiple local
$\chi^2$ minima exist with comparable probabilities. 
Thus, these results are representative of parameters that can
fit the non-zero closure phase and the visibilities, but are {\em not unique}. More short-baseline 
data will be needed to confirm the ``disk hot spot'' model and
to {\em unambiguously} determine the hotspot location.}
\tablenotetext{b}{The predictions of this model are compared to the data in Figures~\ref{v2a_fig}~and~\ref{cp_fig}.}
\end{deluxetable}

\clearpage

\clearpage

\end{document}